\documentclass{PoS}

\title{Moments of meson distribution functions with dynamical twisted mass fermions}

\ShortTitle{Moments of meson distribution functions with dynamical twisted mass fermions}

\author{R$\acute{\mbox{e}}$mi Baron\\
SPhN-DAPNIA, CEA Saclay, 91191 Gif sur Yvette, France\\
E-mail: \email{remi.baron@cea.fr}}

\author{Stefano Capitani\\
Fakult\"at f\"ur Physik, Universit\"at Bielefeld, Universit\"atsstrasse, 
33615 Bielefeld, Germany\\
E-mail: \email{capitani@physik.uni-bielefeld.de}}

\author{Jaume Carbonell\\
Laboratoire de Physique Subatomique et Cosmologie,  53 Av des martyrs, 38026 Grenoble, France\\
E-mail: \email{carbonel@lpsc.in2p3.fr}}

\author{Karl Jansen\\
DESY, Platanenallee 6, 15738 Zeuthen, Germany\\
E-mail: \email{Karl.Jansen@desy.de}}

\author{\speaker{Zhaofeng Liu} and Olivier P$\grave{\mbox{e}}$ne\\
Laboratoire de Physique Th$\acute{e}$orique
(B$\hat{\mbox{a}}$t. 210), Universit$\acute{e}$ de
Paris XI, Centre d'Orsay, 91405 Orsay-Cedex, France\\
E-mail: \email{zhaofeng.liu@th.u-psud.fr},
        \email{Olivier.Pene@th.u-psud.fr}}

\author{Carsten Urbach\\
Theoretical Physics Division, Dept. of Mathematical Sciences,
University of Liverpool, Liverpool L69 7ZL, UK\\
E-mail: \email{Carsten.Urbach@liverpool.ac.uk}}

\author{On behalf of the ETM Collaboration}

\abstract{
We present our preliminary results on the lowest moment 
$\langle x\rangle$ of quark
distribution functions of the pion using two flavor dynamical
simulations with Wilson twisted mass fermions at maximal twist. 
The calculation is
done in a range of pion masses from 300 to 500 MeV. A stochastic
source method is used to reduce inversions in calculating propagators.
Finite volume effects at the lowest quark mass are examined by 
using two different lattice volumes. Our results show that we
achieve statistical errors of only a few percent. We plan to
compute renormalization
constants non-perturbatively 
and extend the calculation to two more lattice spacings and to the nucleons.
}

\FullConference{The XXV International Symposium on Lattice Field Theory\\
		 July 30 - August 4 2007\\
		 Regensburg, Germany}

\begin{document}
\section{Introduction}
Deep inelastic structure functions of mesons and nucleons are
interesting since those functions give us information about the 
momentum and spin
carried by quarks and gluons inside hadrons.
Lattice QCD can calculate moments of structure functions from first principles.
By doing an operator product expansion on the hadronic tensor,
the moments of structure functions are related to
reduced matrix elements of certain local operators. For example, for the
case of scattering of a spin one particle, the details are given
in Ref.\cite{Hoodbhoy:1988am}.
On the lattice, calculations on moments of structure
functions of mesons and nucleons can be found in, e.g., 
Ref.\cite{Gockeler:1995wg,Best:1997qp,Guagnelli:2004ga,Galletly:2005db}.

The Wilson twisted mass formulation of lattice QCD was introduced 
in Ref.\cite{Frezzotti:2000nk,Frezzotti:2003ni}. It provides automatic
$\mathcal{O}(a)$ improvement when tuned to maximal twist, which
can be achieved by setting the PCAC mass to zero. Also,
twisted mass fermions are protected from unphysical fermion zero
modes, thus the problem of exceptional configurations is avoided.
Quenched simulations with Wilson twisted mass fermions have been 
shown to be 
successful\cite{Jansen:2003ir,Jansen:2005gf,Jansen:2005kk,AbdelRehim:2005gz}. 
Specifically, a
quenched calculation
of the lowest moment of the quark distribution function $\langle x\rangle$
in a pion was done in Ref.\cite{Capitani:2005jp}.
Recently, two flavor dynamical simulations\cite{Boucaud:2007uk}
with Wilson twisted mass fermions
gave precise results on low energy constants in the chiral effective
Lagrangian.
Using those configurations, in this work
we calculate
$\langle x\rangle$
in a pion. 
The computation is performed at pion masses in the range
of $\sim300-500$ MeV. In Table~\ref{par}, the parameters of our simulations
are collected. The gauge action used in the simulations is the tree-level
Symanzik improved gauge
action. 
For more details of our data,
please see Ref.\cite{Urbach:lat07}.
\begin{table}[hb]
\begin{center}
\begin{tabular}{ccccccc}
\hline\hline
$\beta$ & $a$(fm) & $a\mu$ & $m_\pi$(GeV) & $L^3\times T$ & $N_{\mbox{meas}}$
& $\langle x\rangle^{\mbox{bare}}$ \\
\hline
3.9& 0.0855(6) & 0.0100 & 0.4839(12) & $24^3\times 48$ & 170 & 0.295(3) \\
& & 0.0085 & 0.4470(12) & $24^3\times 48$ & 167 & 0.279(4) \\
& & 0.0064 & 0.3903(9) & $24^3\times 48$ & 250 & 0.279(5) \\
& & 0.0040 & 0.3131(16) & $24^3\times 48$ & 230 & 0.268(8) \\
& & 0.0040 & 0.3082(55) & $32^3\times 64$ & 316 & 0.251(8) \\
\hline
\hline
\end{tabular}
\caption{Simulation parameters, the number of
measurements($N_{\mbox{meas}}$) and preliminary results of
$\langle x\rangle^{\mbox{bare}}$. Measurements were done
on gauge configurations separated by 20 HMC trajectories
with trajectory length $\tau=1/2$.}
\label{par}
\end{center}
\end{table}

\section{Methodology}
By using the optical theorem, the cross section of deep inelastic
scatterings can be related to the hadronic tensor or the imaginary part
of the forward current-hadron scattering amplitude $W^{\mu\nu}$, which
is given by
\begin{equation}
W^{\mu\nu}(p,q,\lambda,\lambda')=\frac{1}{4\pi}\int d^4x
e^{iq\cdot x}\langle p,\lambda'|[j^\mu(x),j^\nu(0)]|p,\lambda\rangle.
\end{equation}
Here $\lambda$ and $\lambda'$ are the polarization of the target hadron.
$p$ and $q$ are the momenta of the hadron and the virtual photon.
Depending on the spin of the hadron, the above tensor $W^{\mu\nu}$
can be decomposed into a number of independent structure functions.
The spin-averaged structure functions $F_1(x,Q^2)$ and $F_2(x,Q^2)$ 
can tell us overall densities of quarks
and gluons in a hadron. In the parton model, to leading order,
the single flavor structure function $F_1(x)$ is half the probability
of finding a quark with momentum fraction $x$.
If defining the $n$th moment of a function $f(x)$ as
\begin{equation}
M_n(f)=\int_0^1x^{n-1}f(x)dx,
\end{equation}
then for the pion,
to leading twist order, we have from the operator product expansion of
$W^{\mu\nu}$
\begin{equation}
2M_n(F_1)=C_n^{(1)}v_n,\quad M_{n-1}(F_2)=C_n^{(2)}v_n,
\quad (n\ge2, \mbox{even)},
\label{opemn}
\end{equation}
where $C_n^{(k)}=1+\mathcal{O}(\alpha_s)$ are the Wilson coefficients,
and the reduced matrix element $v_n$ is defined by
\begin{equation}
\langle\vec p|\mathcal{O}^{\{\mu_1\cdots\mu_n\}}-\mbox{traces}|
\vec p\rangle=2v_n[p^{\mu_1}\cdots p^{\mu_n}-\mbox{traces}].
\label{vn}
\end{equation}
Here $\{\cdots\}$ means symmetrization on the Lorentz indices and
the twist-2 operators are given by
\begin{equation}
\mathcal{O}^{\mu_1\cdots\mu_n}=(\frac{i}{2})^{n-1}G_{ff'}
\bar\psi_f\gamma^{\mu_1}\stackrel{\leftrightarrow}{D}^{\mu_2}\cdots
\stackrel{\leftrightarrow}{D}^{\mu_n}\psi_{f'},
\end{equation}
where
$\stackrel{\leftrightarrow}{D}=\stackrel{\rightarrow}{D}-
\stackrel{\leftarrow}{D}$
and $G_{ff'}$ is a diagonal flavor matrix.

To compute the lowest moment of the quark distribution function 
$\langle x\rangle$, we need to consider $n=2$ in Eq.(\ref{opemn}).
The corresponding twist-2 operator has
two irreducible representations on the lattice\cite{Gockeler:1996mu}.
In our calculation we use the following operator
\begin{equation}
\mathcal{O}_{44}(x)=\frac{1}{2}\bar u(x)[\gamma_4
\stackrel{\leftrightarrow}{D}_4-\frac{1}{3}\sum_{k=1}^{3}\gamma_k
\stackrel{\leftrightarrow}{D}_k]u(x),
\end{equation}
where
$D_\mu=\frac{1}{2}
(\bigtriangledown_\mu+\bigtriangledown_\mu^*)$ with 
$\bigtriangledown_\mu$ ($\bigtriangledown_\mu^*$) being the usual 
forward (backward) derivative on the lattice.
With the above operator, no external momentum is needed in our 
calculation, which is advantageous since
an external momentum increases the noise to signal ratio. 
One can also use an operator from the other representation, which needs
an external momentum.
In the
continuum limit, the two operators should give the same result
on $\langle x\rangle$. 

From Eq.(\ref{vn}) the bare moment $\langle x\rangle^{\mbox{bare}}$ 
is given by
\begin{equation}
\langle x\rangle^{\mbox{bare}}=v_2=
\frac{1}{2m_\pi^2}\langle\pi,\vec0|\mathcal{O}_{44}
|\pi,\vec0\rangle,
\end{equation}
where the matrix element
$\langle\pi,\vec0|\mathcal{O}_{44}|\pi,\vec0\rangle$ between two
pions at rest is calculated from
the ratio of the following 3-point function and 2-point function with
a source at $t=0$ and a sink at $T/2$:
\begin{equation}
\langle\pi,\vec0|\mathcal{O}_{44}|\pi,\vec0\rangle=
4m_\pi\frac{C_{44}(t)}{C_\pi(T/2)}\quad
(0\ll t\ll T/2).
\end{equation}
Here 
\begin{equation}
C_{44}(t)=\sum_{\vec x\vec y}\langle PS(T/2,\vec x)
\mathcal{O}_{44}(t,\vec y)
PS^\dagger(0,0)\rangle,
\end{equation}
\begin{equation}
C_\pi(T/2)=\sum_{\vec x}\langle PS(T/2,\vec x)PS^\dagger(0,0)\rangle,
\end{equation}
and
$PS(x)=\bar u(x)\gamma_5 d(x)$ is the interpolating field for the pion.
There are two contributions in the Wick contractions of $C_{44}(t)$: one
connected diagram and one disconnected diagram.
The disconnected contribution 
is ignored in our calculation at this moment,
but will be computed by using a stochastic source method.

The above two and three point correlators are evaluated by using a stochastic
time slice source (Z(2)-noise in both real and imaginary part)
\cite{Dong:1993pk,Foster:1998vw,McNeile:2006bz} for all
color, spin and spatial indices. i.e., the quark propagator $X^b_\beta(y)$ is
obtained by solving
\begin{equation}
\sum_yD^{ab}_{\alpha\beta}(z,y)X^b_\beta(y)=\xi(\vec z)^a_\alpha
\delta_{z_0,0}\quad (\mbox{source at}~ t=0),
\end{equation}
where the Z(2) random source $\xi(\vec z)^a_\alpha$ satisfies
the random average condition
\begin{equation}
\langle\xi^*(\vec x)^a_\alpha\xi(\vec y)^b_\beta\rangle=
\delta_{\vec x,\vec y}\delta_{a,b}\delta_{\alpha,\beta}.
\end{equation}
The generalized propagator\cite{Martinelli:1987zd} $\Sigma^b_\beta(y)$
needed in computing
$C_{44}(t)$ is obtained by solving
\begin{equation}
\sum_yD^{ab}_{\alpha\beta}(z,y)\Sigma^b_\beta(y)=\gamma_5X^a_\alpha(z)
\delta_{z_0,T/2}\quad (\mbox{sink at}~ t=T/2).
\end{equation}
The advantage of using the above stochastic source compared with a point
source is that much less inversions are needed.
With a point source, 24 inversions per gauge configuration
are needed: 12 (3 colors $\times$ 4 spins) for the
quark propagator and 12 for the generalized propagator.
With the stochastic source, only two inversions are needed.
One for the
quark propagator, another one for the generalized propagator.
Note that this saving ratio in the number of inversions is special for
the pseudoscalar meson. For other meson correlators, smaller saving ratios
can be achieved.

In our data analysis, measurements are done 
on gauge configurations separated by 20 HMC trajectories
with trajectory length $\tau=1/2$. Statistical errors
are from a Gamma-function
analysis\cite{Wolff:2003sm}.

\section{Preliminary results and outlook}
\begin{figure}
\begin{center}
\includegraphics[width=2in,height=2in]{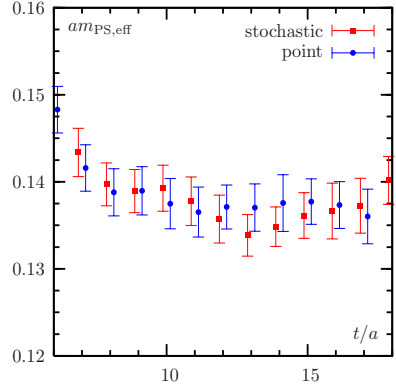}
\end{center}
\caption{Comparison of the pion effective masses for $a\mu=0.004$ 
obtained from a point
source and a stochastic source on the $24^3\times48$ lattice. They are
from the same 230 gauge configurations.}
\label{meffpi}
\end{figure}

\begin{figure}
\begin{center}
\includegraphics[height=65mm,width=65mm]
{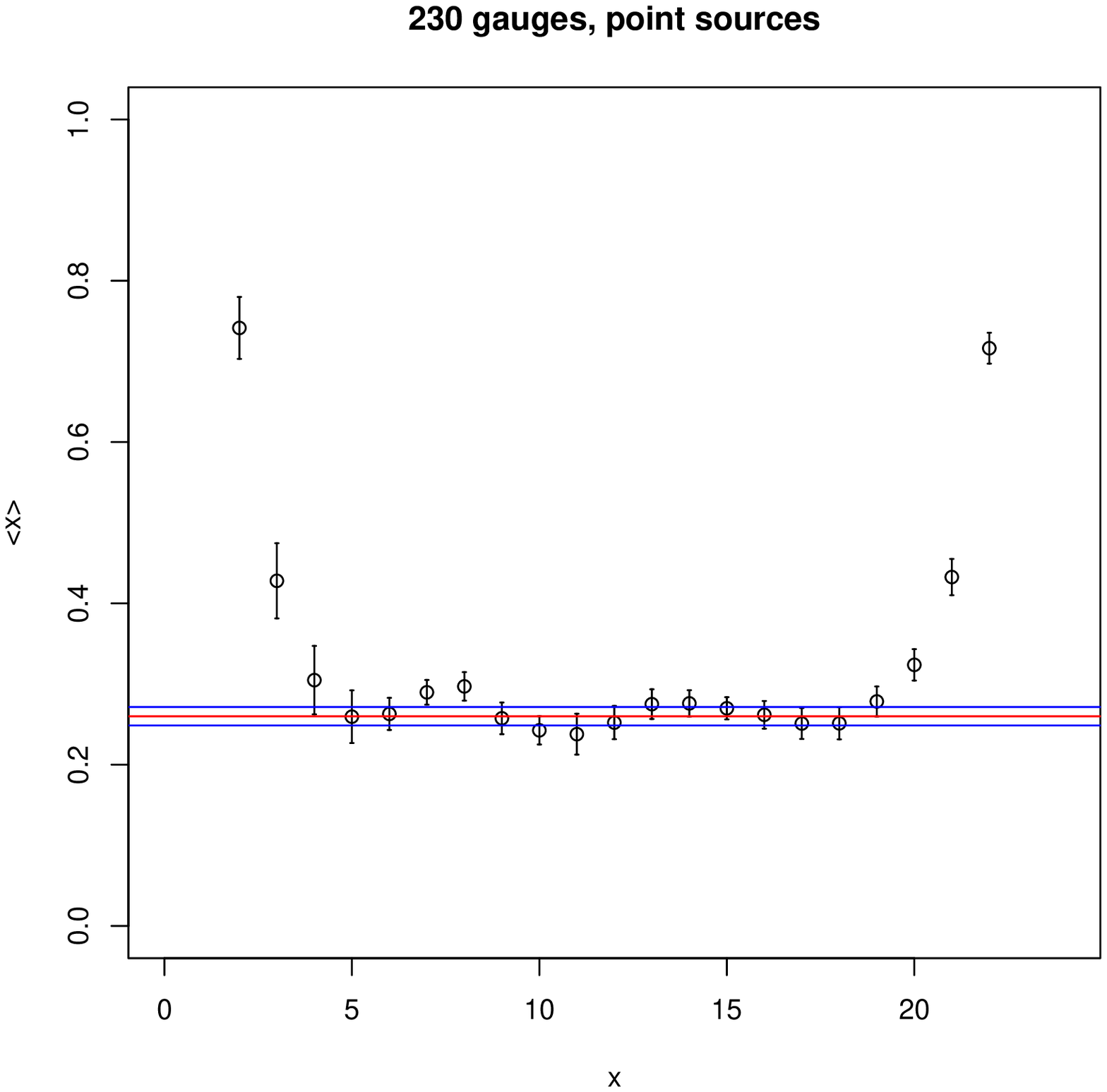}
\includegraphics[height=65mm,width=65mm]
{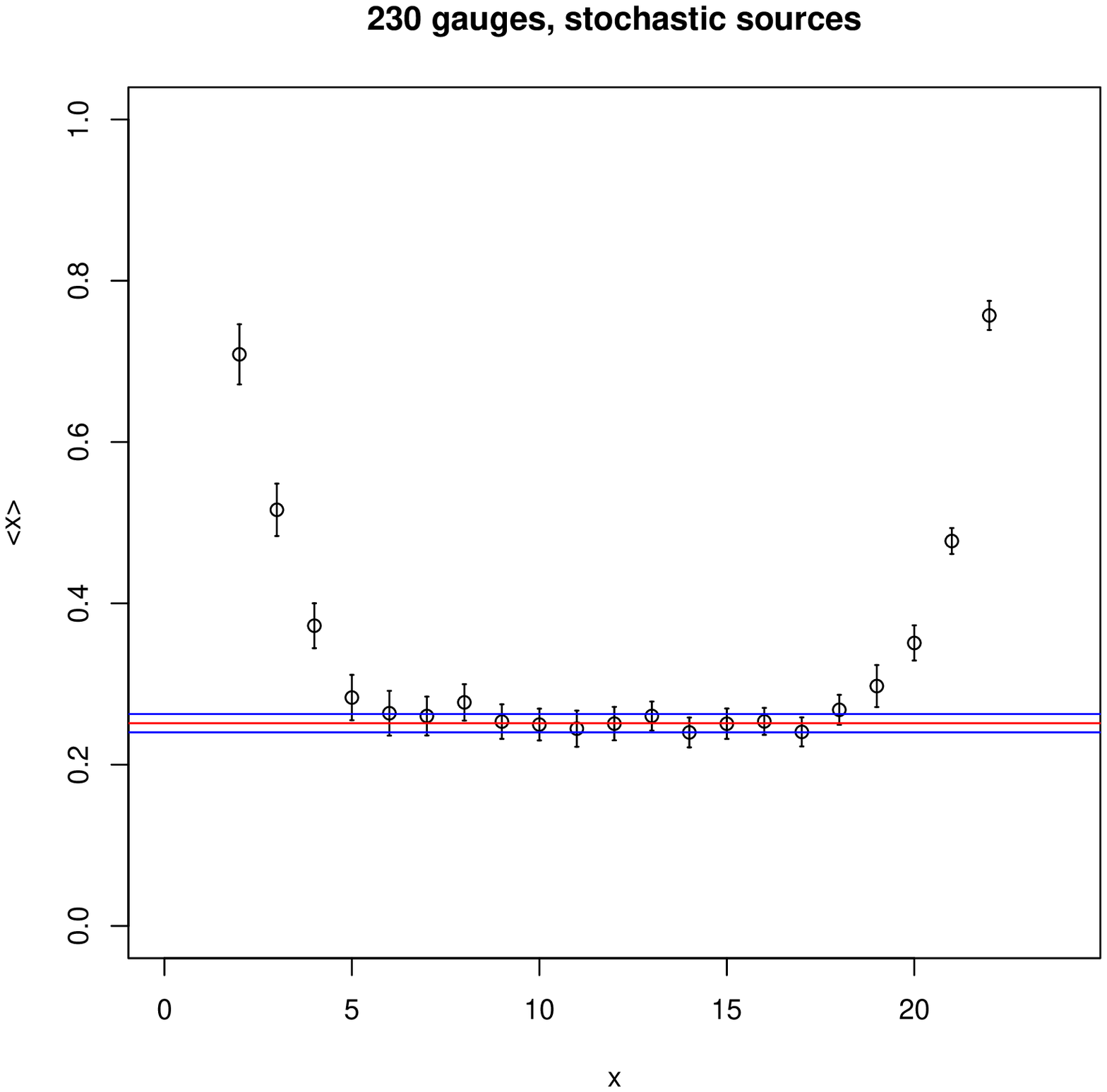}
\end{center}
\caption{$\langle x\rangle^{\mbox{bare}}$ for $a\mu=0.004$ obtained
from a point
source and a stochastic source. We show the average of the two plateaux
around $T/4$ and $3T/4$. The stochastic source method is much cheaper
in computer time.}
\label{x.pt.stoch}
\end{figure}

At the smallest quark mass, $a\mu=0.0040$, we calculated $\langle x\rangle$
using both a point source method and 
the above stochastic
source method on the $24^3\times48$ lattice. 
The results are in agreement within errors.
Fig.\ref{meffpi} shows the comparison of the pion effective
masses obtained from the two methods on
230 gauge configurations. 
As we can see, the statistical errors from the two
methods are of the same size.
In Fig.~\ref{x.pt.stoch}, we compare the results of
$\langle x\rangle^{\mbox{bare}}$ obtained by using the two sources.
The two plateaus of $\langle x\rangle^{\mbox{bare}}$ around $T/4$ and $3T/4$
are averaged to increase statistics.
Again, we find the statistical errors
from the two methods are comparable. However, the stochastic source method
is much cheaper in computer time.

Our preliminary results of $\langle x\rangle^{\mbox{bare}}$ 
are collected in the last column of Table~\ref{par}.
We only give bare quantities here
since the renormalization
constant of the matrix element has not been calculated yet,
which we plan to compute non-perturbatively.
\begin{figure}
\begin{center}
\includegraphics[height=80mm,width=80mm]
{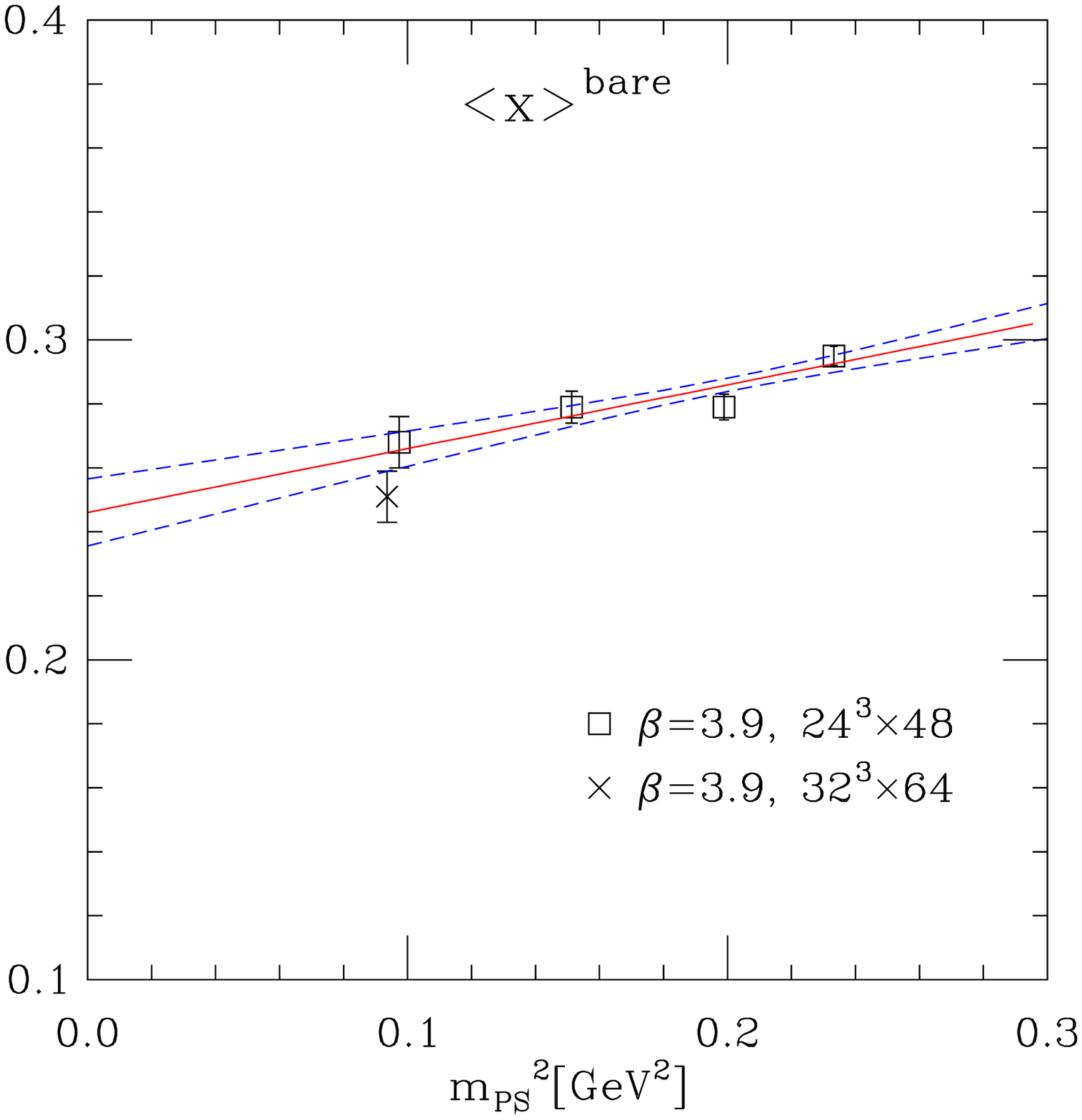}
\end{center}
\caption{$\langle x\rangle^{\mbox{bare}}$ against the pion mass squared.
Finite size effects at the lowest pion mass are not big. 
A linear extrapolation gives $\langle x\rangle^{\mbox{bare}}$=0.246(10) 
in the chiral limit
(using only the points from the $24^3\times48$ lattice).}
\label{xbare}
\end{figure}
Fig.~\ref{xbare} shows the results of $\langle x\rangle^{\mbox{bare}}$
against the pion mass
squared. The cross in the graph at the 
lowest quark mass is from the $32^3\times64$ lattice.
It shows that the finite lattice volume effects are not big.
A linear extrapolation in $m_\pi^2$ gives
$\langle x\rangle^{\mbox{bare}}$=0.246(10) in the chiral limit
(using only the points from the $24^3\times48$ lattice).

Using Wilson twisted mass fermions, our two flavor dynamical simulations
go down to a pion mass of around 300 MeV. Our first attempt in this work
shows we can get $\langle x\rangle$ with small statistical errors of
a few percent.
This will enable us to compare numerical results of $\langle x\rangle$ 
in pions and nucleons with predictions
from the chiral perturbation theory (see for example 
\cite{Arndt:2001ye,Chen:2001eg,Detmold:2005pt}). 
With simulations at two more lattice spacings,
we will be able to do an
extrapolation to the continuum
limit (the data analysis is in progress). 
For the future, we plan to compute the disconnected diagrams using a stochastic
source method and calculate the renormalization constant non-perturbatively.
We also plan to compute $\langle x\rangle$ in nucleons.

\section*{Acknowledgments}
The numerical calculations were performed on the ZOO cluster at LPT in Orsay,
the computers of the CCRT (Bruy$\grave{\mbox{e}}$re-le-Chatel) computing center
and the Blue Gene/L in J{\"u}lich. We thank Chris Michael for discussions
on the stochastic source method.
Zhaofeng Liu thanks Stefan Schaefer and Andrea Shindler for 
valuable discussions.

\end{document}